\documentclass[bibyear]{aa} % if the references are not structured
\usepackage{graphicx}
\usepackage{txfonts}                 %  Postscript Fonts
\newcommand{\kms}{km\,s$^{-1}$}

\newcommand{\xiA}{$\xi$\,Boo\,A}
\newcommand{\xiB}{$\xi$\,Boo\,B}

\begin{document}

\titlerunning{ZDI of $\xi$\,Boo\,A+B }
\authorrunning{Strassmeier, Carroll \& Ilyin}

\title{Zeeman Doppler Imaging of $\xi$\,Boo\,A and B\thanks{Based on data acquired with the Potsdam Echelle Polarimetric and Spectroscopic Instrument (PEPSI) using the Large Binocular Telescope (LBT). The LBT is an international collaboration among institutions in the United States, Italy, and Germany.}}

   \author{K.~G.~Strassmeier\inst{1}
        \and T.~A.~Carroll\inst{1}
        \and I.~V.~Ilyin\inst{1}
   }

   \institute{Leibniz-Institute for Astrophysics Potsdam (AIP),
     An der Sternwarte 16, 14482 Potsdam, Germany\\
     \email{kstrassmeier@aip.de,tcarroll@aip.de,ilyin@aip.de}
}

   \date{Received xxx x, 2022; accepted xxx x, 3022}

  \abstract
   {}
   {We present a magnetic-field surface map for both stellar components of the young visual binary $\xi$\,Boo~AB (A: G8V, B: K5V). }
   {Employed are high resolution Stokes-V spectra obtained with the Potsdam Echelle Polarimetric and Spectroscopic Instrument (PEPSI) at the Large Binocular Telescope (LBT). Stokes~V line profiles are inverted with our $i$MAP software and compared to previous inversions. We employed an iterative regularization scheme without the need of a penalty function and incorporated a three-component description of the surface magnetic-field vector. The spectral resolution of our data is 130\,000 (0.040--0.055\,\AA ) and have signal-to-noise ratios (S/N) of up to three thousand per pixel depending on wavelength. A singular-value decomposition (SVD) of a total of 1811 spectral lines is employed for averaging Stokes-V profiles. Our mapping is accompanied by a residual bootstrap error analysis.}
   {Magnetic flux densities of the radial field component of up to plus/minus 115$\pm$5\,G were reconstructed for \xiA\ while up to plus/minus 55$\pm$3\,G were reconstructed for \xiB. \xiA's magnetic morphology is characterized by a very high latitude, nearly polar, spot of negative polarity and three low-to-mid latitude spots of positive polarity while \xiB's morphology is characterized by four low-to-mid latitude spots of mixed polarity. No polar magnetic field is reconstructed for the cooler \xiB\ star. Both our maps are dominated by the radial field component, containing 86\%\ and 89\%\ of the magnetic energy of $\xi$\,Boo A and B, respectively. We found only weak azimuthal and meridional field densities on both stars (plus/minus 15--30~G), about a factor two weaker than what was seen previously for \xiA. The phase averaged longitudinal field component and dispersion is +4.5$\pm$1.5\,G for \xiA\ and --5.0$\pm$3.0\,G for \xiB. }
     {}

   \maketitle

%-------------------------------------------------------------------
\section{Introduction}

The technique of Zeeman-Doppler imaging (ZDI) was introduced three decades ago by Semel (\cite{semel}) and had become an indispensable technique for cool-star magnetic field studies (see reviews by Kochukhov \cite{koch}, Reiners \cite{rein}, Strassmeier \cite{spots}, and Donati \& Landstreet \cite{don:lan}). So far, ZDI employed almost exclusively circular polarization (CP) Stokes~V spectra and only rarely also Stokes~IV, that is, intensity and CP line profiles together. This is because Stokes~I inversion is not feasible for stars with low rotational line broadening due to the Doppler effect of the surface rotation being too small. The targets $\xi$~Boo A+B are among those cool stars with very small rotational line broadening but still with an active atmosphere. Therefore, only CP ZDI of the $\xi$\,Boo AB binary system is possible.

$\xi$~Boo's optical light variability was discovered by Chugainov (\cite{chu}). He obtained a photometric period of $\approx$10\,d with an average amplitude of just 5.5\,mmag from 45 nights of $UBV$ data in 1980. The variability was assigned to \xiA\ although only the combined light with \xiB\ could be measured. On two nights Chugainov (\cite{chu}) noted flare-like events with $\Delta U$ of up to 0\fm1. Other determinations of the rotational period followed. For \xiA , Plachinda \& Tarasova (\cite{pla:tar}) obtained 6.1455$\pm$0.0003\,d using longitudinal magnetic field measurements. Activity indices based on the \ion{Ca}{ii} H\&K lines were used by Noyes et al. (\cite{noyes}) to obtain 6.2$\pm$0.1\,d, by Donahue et al. (\cite{don:saa}) to obtain 6.31\,d, and by Hempelmann et al. (\cite{hemp}) to obtain 6.299$\pm$0.037\,d. Toner \& Gray (\cite{ton:gra}) employed spectral line bisectors and line ratios to get 6.43$\pm$0.01\,d.  As suggested by Morgenthaler et al. (\cite{morgen}) the differences between these values may be related to their probing of different stellar layers and latitudes and the star's differential rotation. For \xiB , a mean period of 11.94\,d was obtained by Donahue et al. (\cite{don:saa}) from the Mt.~Wilson \ion{Ca}{ii} data. Its period range with a minimum value of 10.92\,d and a maximum of 13.19\,d was also attributed to differential surface rotation. Vidotto et al. (\cite{vid}) listed 10.3\,d for \xiB\ in their sample table and referred to a paper by Petit et al. in preparation, the origin of this period is thus unclear.
%\footnote{we note that in the paper by Baliunas et al. (\cite{bal}) she just referred to a period of 11\,d}

A new determination of the lithium abundance of both $\xi$\,Boo stars was recently presented by Strassmeier \& Steffen (\cite{str:ste}), who also reviewed the  relevant global stellar parameters of both components together with the K5 reference dwarf star 61\,Cyg\,A. We adopt their astrophysical parameters for our ZDI input in the present paper but also refer to Takeda et al. (\cite{takeda}) for an independent determination of the stellar parameters of the \xiA\ component. The $^7$Li abundance is 23 times higher for \xiA\ than the Sun's, but three times lower than the Sun's for \xiB\ while both fit the trend of single stars in the similar-aged M35 open cluster. Both star's age appears to be $\approx$200\,Myr.

Magnetic-field measurements concentrated mostly on the brighter A component. Boesgaard (\cite{bos74}) presented an initial marginal detection based on a conventional Zeeman analyzer while Boesgaard et al. (\cite{bos75}) followed up and concluded that the field was too weak for their kind of technique. The A component's surface magnetic field was firstly and conclusively detected from line broadening by Robinson et al. (\cite{rob80}) and was of appreciable strength of up to 2.9\,kG. It was followed-up with a surprising null detection by Marcy (\cite{m81}). Later, Marcy (\cite{m84}) confirmed the field, albeit less strong, from further Zeeman broadening data and even found a 670\,G field and a filling factor of 73\% on the B component. Infrared measurements of Zeeman sensitive Ti lines allowed Saar (\cite{s94}) to measure 2.3\,kG and a filling factor of 20\% for the B component. With the same technique, but based on four infrared Fe lines, Gondoin et al. (\cite{gon}) reported another non-detection for \xiA. The discrepant field strengths and sometimes even null detections for the brighter A component were suggested to be due to intrinsic variability. The rotational modulation of the large-scale field was then first investigated by Plachinda \& Tarasova (\cite{pla:tar}) from a collection of longitudinal field measurements collected over two decades. Using observations mostly from 1998, along with archival polarization measurements from Borra et al. (\cite{borra}) and Hubrig et al. (\cite{hub}), Plachinda \& Tarasova (\cite{pla:tar}) even reported a sign reversal of the longitudinal field. From higher resolution observations, Petit et al. (2005) found that \xiA\ has a magnetic field made up of two morphological components; a 40\,G dipole inclined at 35\degr\ to the rotation axis, and a large-scale 120\,G toroidal field. The BCool snapshot survey (Marsden et al. \cite{marsden}), also based on Stokes~V spectra, claimed longitudinal fields of between 18.4$\pm$0.3\,G and 0.5$\pm$1.0\,G for the A component and an equally strong but negative field of --18.9$\pm$0.5\,G for the B component. The first ZDI maps of \xiA , for seven epochs, came from Morgenthaler et al. (\cite{morgen}). They followed up on the morphology suggestion from Petit et al. (\cite{petit}) and found that the toroidal component persists with a constant polarity containing a significant fraction of the magnetic energy of the large-scale surface field throughout all observing epochs. The evolution of the field geometry was modeled with an increase of field strength and dipole inclination. The discrepancy of magnetic field strengths from Stokes-V spectropolarimetry and line-broadening studies was addressed again by Kochukhov et al. (\cite{koch20}). They determined field strength and filling factor for \xiA\ from the broadening of specific lines to 1.2$^{+0.4}_{-0.3}$\,kG and 69$\pm$28\,\%, respectively, as compared to 36$^{+26}_{-13}$\,G from ZDI, both based on the data from Morgenthaler et al. (\cite{morgen}).

Cotton et al. (\cite{cott}) had observed \xiA\ with high precision broad-band linear polarimetry contemporaneously with circular spectropolarimetry and confirmed a modulation with a period of 6.43\,d from both techniques. However, the signals from the two techniques were out of phase by 0.25, which the authors interpreted to be due to differential saturation of spectral lines in the global transverse magnetic field. Already in an early study, Huovelin et al. (\cite{huo}) concluded that \xiA\ varied in linear polarization (LP) along its rotational cycle, then based on data that just allowed a 2--2.5$\sigma$ detection from the mean. They also accounted the linear polarization to the resulting saturation in the Zeeman-sensitive absorption lines.

The present paper employs high-resolution data from PEPSI at the 2$\times$8.4\,m LBT. The spectral resolution of PEPSI in polarimetric mode is 130\,000 and is thus twice as high as for previously published maps of \xiA. Moreover, the light-gathering power of the LBT enabled an unprecedented S/N approaching the data quality of solar observations. These new data together with the characteristics of our ZDI code $i$MAP are described in Sect.~\ref{S2}. The Zeeman-Doppler images from Stokes~V are presented in Sect.~\ref{S3}. Section~\ref{S4} summarizes our findings.

%------------------------F1
   \begin{figure*}
   \includegraphics[angle=-90,width=\textwidth, clip]{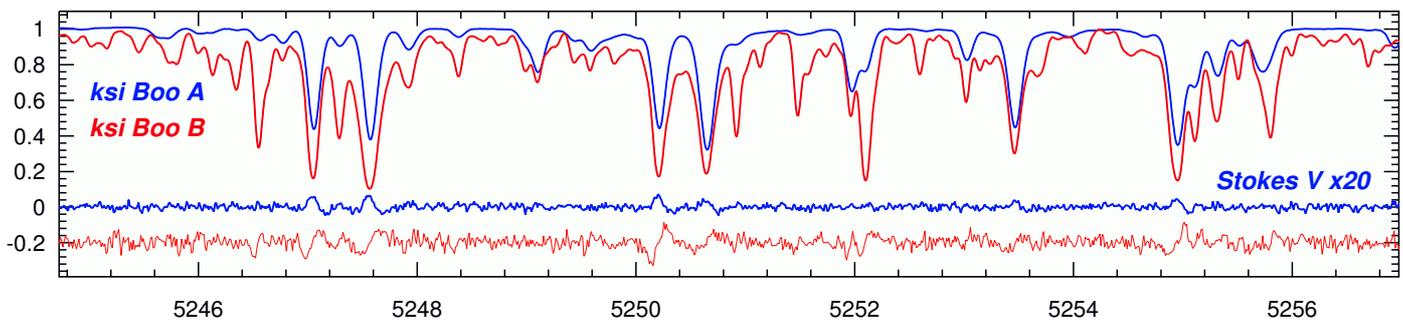}
   \caption{Stokes IV example spectra for a wavelength range of 10\,\AA\ near 5250\,\AA . The vertical axis is relative intensity, the horizontal axis is wavelength in Angstroem.
   Top: Stokes-I spectra of \xiA\ (blue) and \xiB\ (red). Bottom: Stokes-V spectra of \xiA\ (blue) and \xiB\ (red), the latter offset by --0.2 in intensity for better display, and both enhanced in scale by a factor 20 compared to Stokes~I.}
         \label{F1}
   \end{figure*}

%--------------------------------------------------------------------
\section{Observations and ZDI data input}\label{S2}

\subsection{Spectroscopic data}\label{S21}

High-resolution spectra were obtained with PEPSI at the effective 11.8\,m LBT in southern Arizona. We employed both polarimeters in the LBT's two symmetric straight-through Gregorian foci. Two pairs of octagonal 200\,$\mu$m fibers per polarimeter feed the ordinary and extraordinary polarized beams via a five-slice image slicer per fiber into the spectrograph. It produces four spectra per \'echelle order recorded in a single exposure with a spectral resolution of $\lambda/\Delta\lambda$=130\,000 sampled by 4.2 pixels on the CCDs. Each fiber diaphragm appears on the sky as a circular projection with a diameter of 1.5\arcsec , thus easily isolate the 7\arcsec\ separated $\xi$\,Boo binary components. The two polarimeters are identical in design and construction but are separately calibrated. Both are of a classical dual-beam design with a modified Foster prism as linear polarizer with two orthogonally polarized beams exiting in parallel. The achromatic PMMA quarter-wave retarder is located in front of the Foster prism on a rotary stage. The Foster prism, the atmospheric dispersion correctors, two fiber heads, and two fiber viewing cameras are rotating as a single unit with respect to the parallactic axis on sky. This design avoids cross-talk between circular and linear polarization (Ilyin \cite{ilya12}). The spectrograph and the polarimeters were described in detail in Strassmeier et al. (\cite{pepsi}, \cite{spie-austin}).

Observations of both $\xi$\,Boo stars commenced over 10 consecutive nights May 6-16, 2019. Eight and six IQUV\footnote{Analysis of the Stokes QU data is not part of the present paper.} spectra for \xiA\ and \xiB\, respectively, were obtained with cross disperser (CD) III covering 4800--5441~\AA\ and with CD~V covering 6278--7419~\AA . Retarder angles of 45\degr\ and 135\degr\ were set for Stokes~V (with the Foster prism position angle set to 0\degr ). The zero level of the position angle of the Foster prism is aligned to the north direction. Its alignment accuracy is $\approx$1\degr\ or better. Exposure time per integration was 5\,min for \xiA\ and 10\,min for \xiB . For Stokes~V, it resulted in a (quantile 95\%) S/N per pixel of up to 2100 in CD~V and 1700 in CD~III for \xiA , and 1350 and 950 for \xiB, respectively. S/N of CP spectra is in the present case $\approx$60\,\%\ of Stokes~I because a Stokes-I spectrum combines the six QU\&V sub-exposures while V combines only two sub-exposures. At this S/N the polarization line signatures are even recognized by eye as shown in Fig.~\ref{F1} for a wavelength region around 5250\,\AA. The example spectra in this figure show three line pairs with strong Zeeman modulation; \ion{Fe}{i} 5247.1\,\AA , \ion{Cr}{i} 5247.6\,\AA , \ion{Fe}{i} 5250.2\,\AA , \ion{Fe}{i} 5250.7\,\AA , \ion{Fe}{i} 5253.5\,\AA , \ion{Fe}{i} 5254.9\,\AA .  The log of all observations is given in the Appendix in Table~\ref{TA1}.

% ------------------------------ Table 1
\begin{table}
\caption{Adopted astrophysical properties of \xiA\ and \xiB . } \label{T1}
\begin{tabular}{lllll}
\hline \noalign{\smallskip}
Parameter                   & \xiA  & Ref. & \xiB   & Ref.  \\
\noalign{\smallskip}\hline \noalign{\smallskip}
Classification, MK          & G8V          & (1) & K5V          & (1) \\
Effective temperature, K    & 5480         & (2) & 4570         & (3) \\
Log gravity, cgs            & 4.53         & (2) & 5.0          & (2)\\
$v\sin i$, \kms             & 3.0          & (3,4) & 1.5        & (3) \\
Microturbulence, \kms       & 1.4          & (2) & 0.15         & (2) \\
Macroturbulence, \kms       & 3.6          & (5) & \dots        &  \\
Rotation period, d          & 6.43         & (6) & 11.94        & (7) \\
Inclination, deg            & 28$\pm$5     & (8,9)& $\approx$32 & (3) \\
Log metallicity, solar      & --0.13       & (3) & +0.13        & (3) \\
\noalign{\smallskip}\hline
\end{tabular}
\tablefoot{(1) Abt (\cite{abt}). (2) Luck (\cite{luck}). (3) Strassmeier \& Steffen (\cite{str:ste}). (4) Gray (\cite{g84}). (5) Allende Prieto et al. (\cite{allpri}). (6) Toner \& Gray (\cite{ton:gra}). (7) Donahue et al. (\cite{don:saa}). (8) Petit et al. (\cite{petit}). (9) Morgenthaler et al. (\cite{morgen}). }
\end{table}

Data reduction was performed with the software package SDS4PEPSI (``Spectroscopic Data Systems for PEPSI'') based on Ilyin (\cite{4A}), and described in some detail in Strassmeier et al. (\cite{sun}, \cite{pepsi}). The specific steps of image processing include bias subtraction and variance estimation of the source images, super-master flat field correction for the CCD spatial noise, scattered light subtraction, definition of \'echelle orders, wavelength solution for the ThAr images, optimal extraction of image slicers and cosmic spikes elimination, normalization to the master flat field spectrum to remove CCD fringes and the blaze function, a global 2D fit to the continuum, and the rectification of all spectral orders into a 1D spectrum.

%------------------------F2   ksi Boo A
   \begin{figure}
   {\bf a. \hspace{40mm} b.}\\
   \includegraphics[width=43mm,clip]{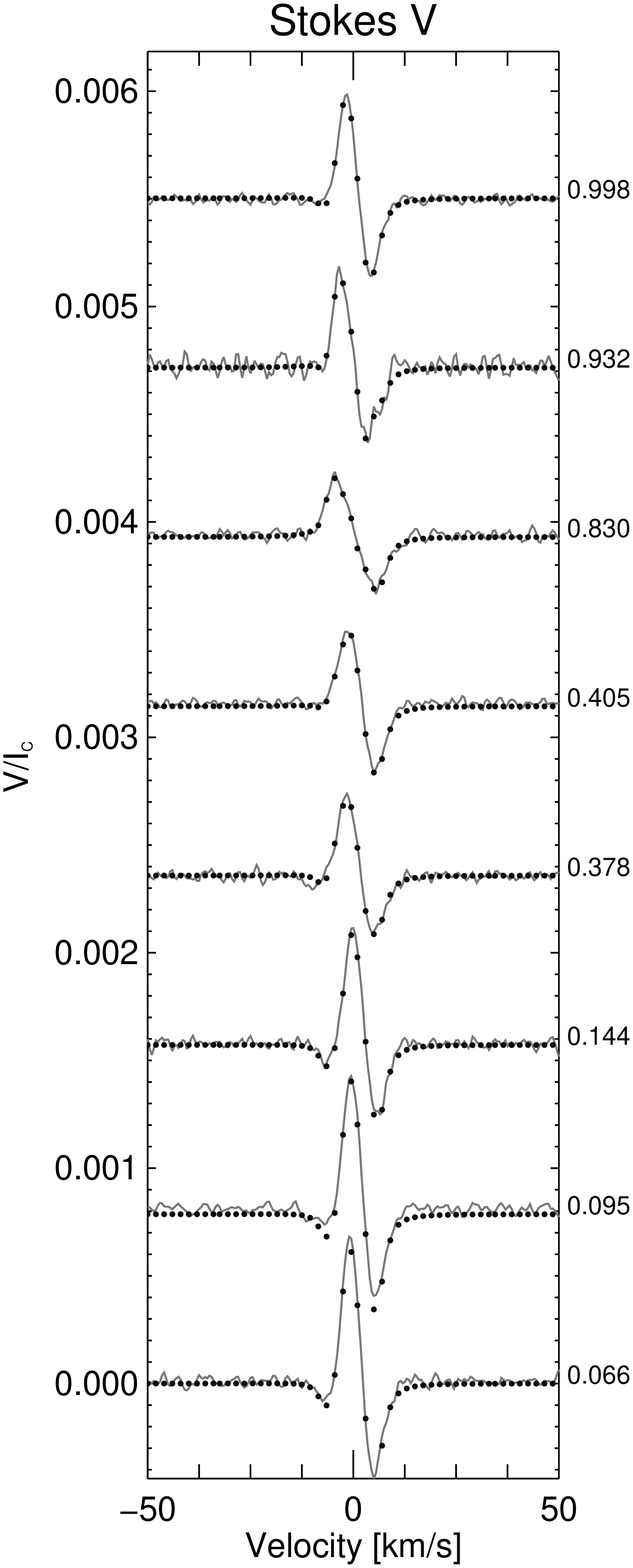}
   \includegraphics[width=43mm,clip]{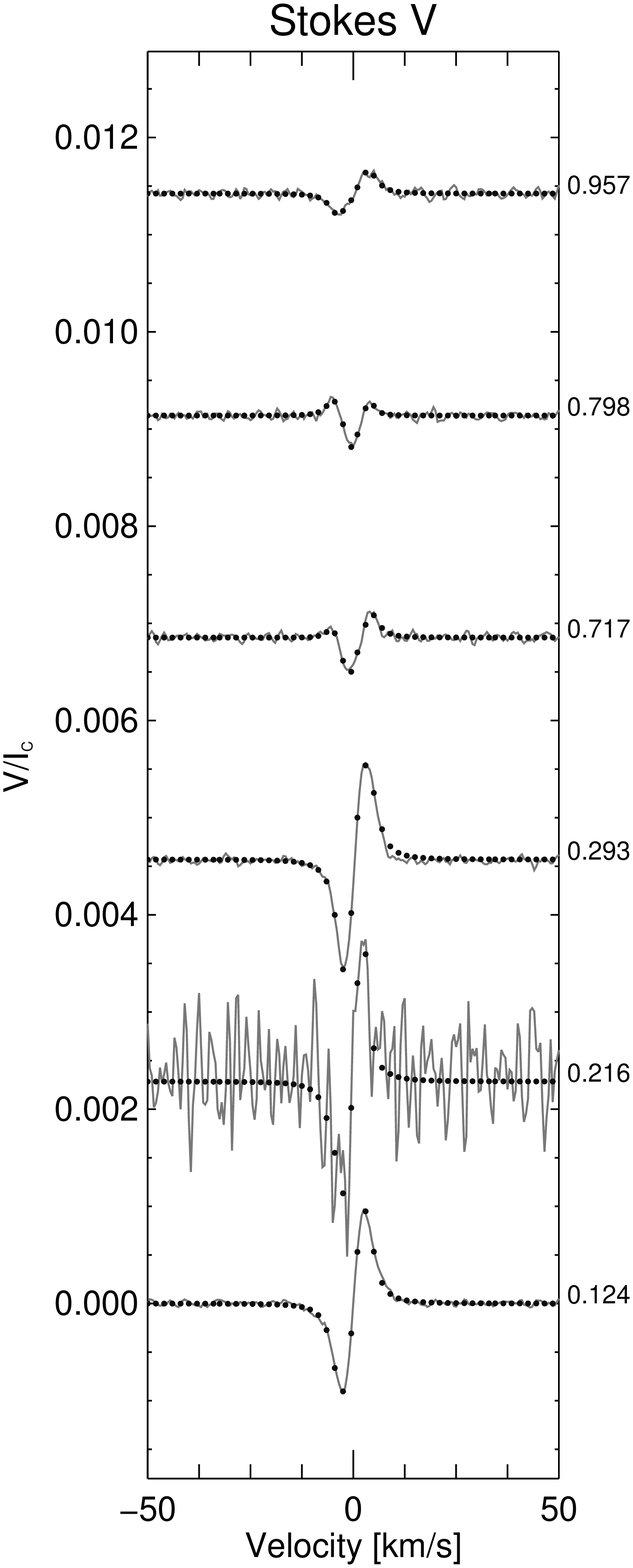}
   \caption{Observed (lines) and inverted (dots) SVD Stokes-V line profiles. Panel $a$: \xiA . Panel $b$: \xiB . Profiles are shifted in intensity arbitrarily and labeled with their respective phases $\phi$ on the right side of each panel. Rotation advances from bottom to top.}
         \label{F2}
   \end{figure}

\subsection{Characteristics of the inversion code $i$MAP}\label{S22}

All image reconstructions in this paper are done with the $i$MAP code (Carroll et al. \cite{carr07}, \cite{carr12}). $i$MAP employs a three-component magnetic-field vector (radial, meridional, azimuthal) per surface pixel instead of the widely used spherical harmonics expansion for its description. We also use an iterative regularization where the step size and an appropriate stopping rule provides the regularization of the inverse problem. Our present inversion technique is thus penalty free and is based on the Landweber iteration to minimize the sum of the squared errors (for more details see Carroll et al. \cite{carr12} and references therein). The code can either perform multi-line inversions for a large number of photospheric line profiles simultaneously or use a single average SVD-extracted line profile. For the application in this paper, we use the latter for Stokes~V. Its eigenvalue decomposition of the signal covariance matrix, that is a SVD of the observation matrix, emphasizes the similarity of the individual Stokes profiles and allows one to identify the most coherent and systematic features. Incoherent features, like noise and line blends, will be dispersed along many dimensions in the transformed eigenspace.

The stellar surface is partitioned into 5\degr~$\times$~5\degr\ segments, resulting in 2592 surface pixels for the entire sphere. The local line-profile computation in $i$MAP is based on a radiative transfer solution with the help of an artificial neural network (Carroll et al.~\cite{carr08}). The atomic parameters for the line synthesis are taken from the Vienna Atomic Line Database VALD3 (e.g., Ryabchikova et al.~\cite{vald}). These are used with a grid of Kurucz ATLAS-9 model atmospheres (Castelli \& Kurucz~\cite{atlas-9}) for local line profiles in 1D and in LTE. The grid covers temperatures between 3500\,K and 8000\,K in steps of 250\,K interpolated to the gravity and microturbulence values from Table~\ref{T1}. Table~\ref{T1} is a quick-look version of the ZDI relevant astrophysical parameters of both stars and their adopted values.

%------------------------F3
   \begin{figure*}
   {\bf a.} \ \xiA\\
   \includegraphics[width=\textwidth,clip]{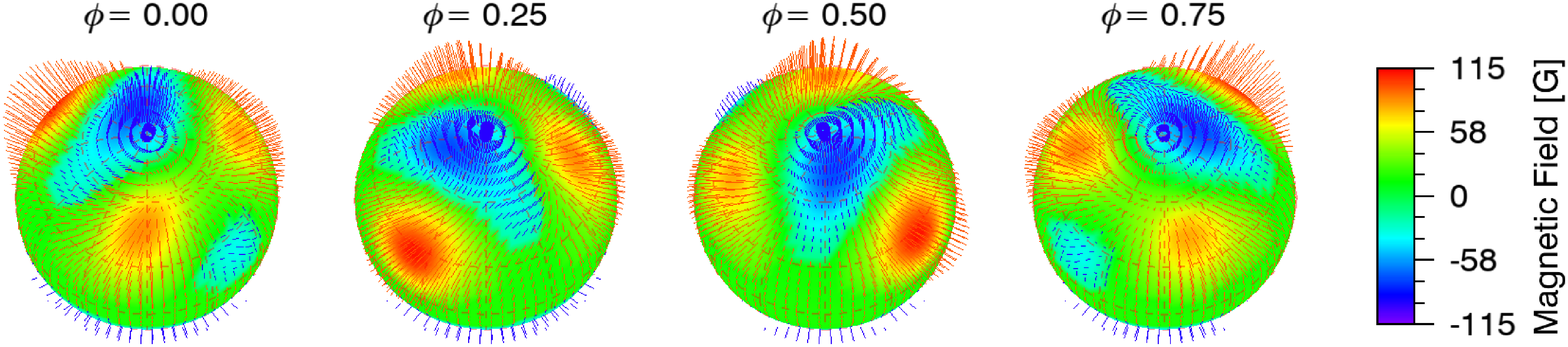}
   {\bf b.} \ \xiB\\
   \includegraphics[width=\textwidth,clip]{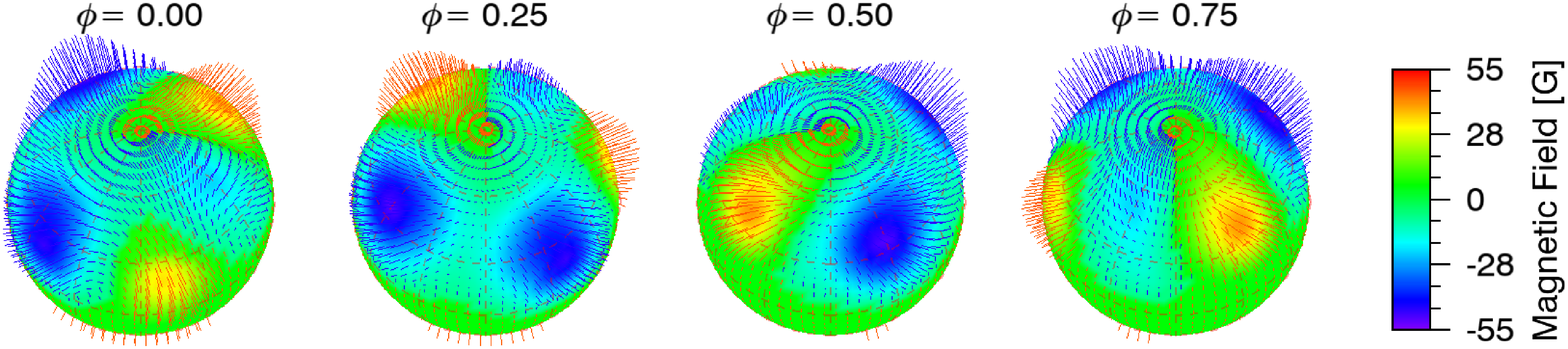}
   \caption{Zeeman-Doppler images of \xiA\ (panel $a$) and \xiB\ (panel $b$) in orthographic projection. $\phi$ is the rotational phase. Magnetic field strength is color coded and identified in the right bars in units of Gauss. Positive polarity is depicted in red, negative polarity in blue. The length of the surface dashes is proportional to field strength. }
         \label{F3}
   \end{figure*}

\subsection{ZDI input}\label{S23}

We created SVD-averaged Stokes-V line profiles from 1811 individual spectral lines. Each of the 1811 lines were modeled and synthesized to produce a single synthetic SVD profile that can then be compared to the observed SVD profile. Instead of building a large database of SVD profiles for all possible field configurations and viewing angles our approach uses an artificial neural network to allow a quasi on-the-fly calculation during the course of the inversion. A detailed description and error analysis of this method is given in Carroll et al. (\cite{carr08}).  The 1811 lines were selected upon line depths larger than 10\%\ with respect to the local continuum and as blend free as possible (following Strassmeier et al. \cite{iipeg}). The average wavelength of the full line list is $\lambda_0=5068$\,\AA\ and the average Land\'e factor $g_0=1.21$. The individual Stokes-V profiles of \xiA\ have S/N of between 1000 -- 2000 per pixel (listed in detail in Table~\ref{TA1}) but tens of thousands per velocity bin per SVD-averaged profile. The respective S/N values for \xiB\ are approximately a factor two lower than for \xiA. A total of eight rotational phases are available for the inversion for \xiA\ and six phases for \xiB.

Rotational phases $\phi$ were calculated from the respective rotation periods and the arbitrary zero point already used by Petit et al. (\cite{petit})  and Morgenthaler et al. (\cite{morgen}) for \xiA ,
\begin{eqnarray} \label{eq1}
\xi\ \mathrm{Boo\ A} \ : \ \ 2,452,817.41  +  6.43 \times\ E \ \phi , \\
\xi\ \mathrm{Boo\ B} \ : \ \ 2,452,817.41  +  11.94 \times\ E \ \phi .
\end{eqnarray}
Our time coverage and phase sampling is too sparse to support a new and better period or differential-rotation determination. Our data were taken within one and a half stellar rotations of \xiA, and basically one rotation of \xiB, and thus do not provide the time span to show the smearing effect from differential surface rotation. As all previous authors, we prefer the 6.43\,d period from Toner \& Gray (\cite{ton:gra}) for \xiA. The 11.94-d period for \xiB\ from Donahue et al. (\cite{don:saa}) from the Mt.~Wilson \ion{Ca}{ii} S-index time series is the only documented measured value in the literature anyway.

Both $\xi$\,Boo stars appear with comparably narrow spectral lines. Its $v\sin i$ values of 3~\kms\ and 1.5~\kms, respectively, are considerably lower than for typical targets employed for Doppler imaging. At a resolving power of 130\,000 (2.3~\kms\ at 6000\,\AA ), and an average full width of the lines at continuum level of $2 \ (\lambda/c) \ v\sin i \approx 0.12$~\AA\ for \xiA, we have only $2\frac{1}{2}$ resolution elements across the rotating stellar disk. According to the Stokes-I simulations of Piskunov \& Wehlau (\cite{pis:weh}) five resolution elements are the minimum for successful Doppler imaging because their test inversions with artificial data showed a reasonably correct recovery of the input image only when at least five resolution elements were available but not with less. We note that even our $2\frac{1}{2}$ resolution elements are twice as many as compared to previous ZDI maps of \xiA\ based on the $R\approx$65\,000 NARVAL spectropolarimeter at the Telescope Bernard Lyot.

%------------------------F4
   \begin{figure*}
   {\bf a.} \ \xiA\\
   \includegraphics[width=\textwidth,clip]{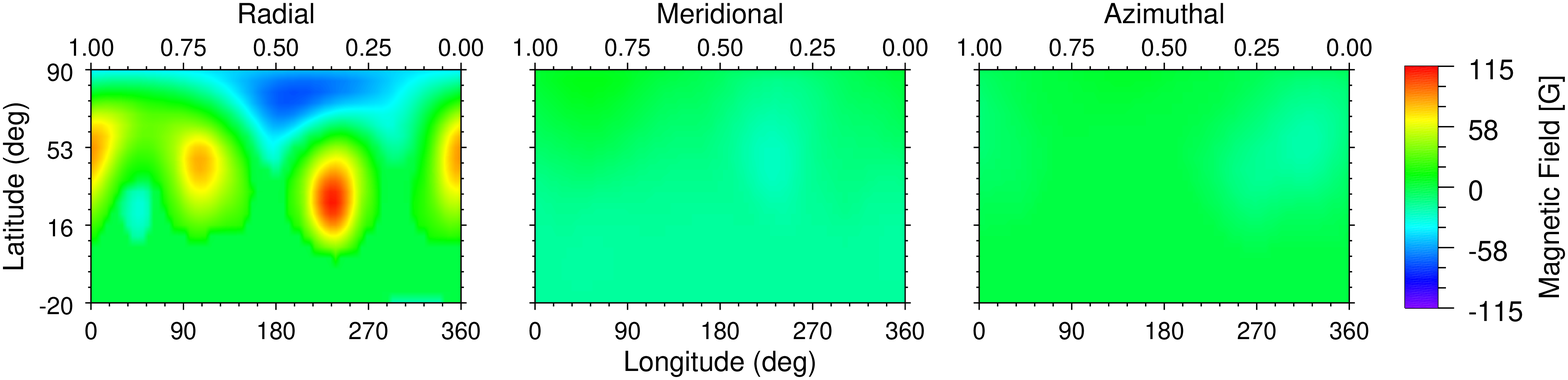}
   {\bf b.} \ \xiB\\
   \includegraphics[width=\textwidth,clip]{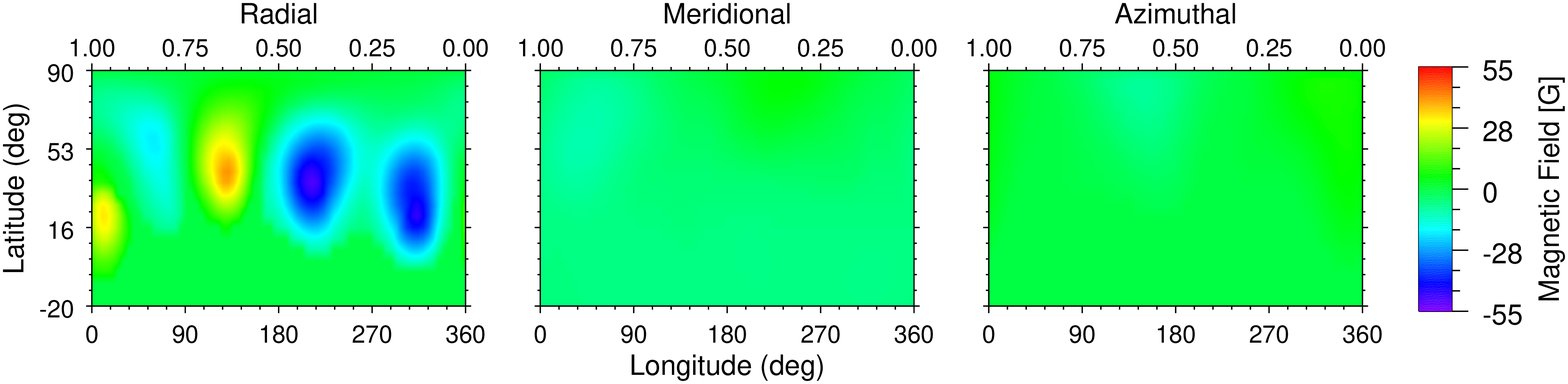}
   \caption{Zeeman-Doppler images in Mercator projection. Panel $a$: \xiA\ . Panel $b$: \xiB . Otherwise as in Fig.~\ref{F3} but split into the three vector components. From left to right: radial-, meridional-, and azimuthal field component in Gauss. }
         \label{F4}
   \end{figure*}

%--------------------------------------------------------------------
\section{Analysis}\label{S3}

\subsection{Stokes V line-profile amplitudes}\label{S31}

While for cool stars the strength of the signal in CP spectral lines rarely exceeds 1\%\ of the continuum intensity and requires truly high S/N to be recognized by eye, the approximately ten-times weaker LP signal (at least for the Sun) is expected to always remain below the noise level. For example, Kochukhov et al.~(\cite{koc:man}) finds a Stokes~V signature at the 2--3$\sigma$ confidence level in only a few of the strongest spectral lines of the super-active star II~Peg in SOFIN data from the 2.4m NOT as well as in ESPaDOnS data from the 3.6m CFHT. Our PEPSI data from the 11.8m LBT had shown it on the 10$\sigma$ level (Strassmeier et al. \cite{iipeg}). The PEPSI spectra for the two comparably low-activity $\xi$\,Boo components in the present paper indicate Stokes~V signatures at the 3$\sigma$ level in the direct spectra, which are easily visible by eye in Fig.~\ref{F1}. The SVD-profiles for \xiA\ show full amplitudes of up to $\approx$0.001 in Stokes~V which is, as expected, significantly smaller (3$\times$) when compared with the super-active star II~Peg despite a $v\sin i$ difference of the two stars of a factor seven.

We calculate the mean longitudinal magnetic field, $\langle B_\mathrm{z} \rangle$ in Gauss, from the first moment of Stokes~V as formulated in Kochukhov et al. (\cite{koch10});
\begin{equation}\label{Bz}
    \langle B_\mathrm{z} \rangle = -2.14 \times 10^{11} \frac{\int v V(v) \,\mathrm{d}v }{\lambda_0 \ g_0 \ \mathrm{c} \int[I_c - I(v)] \,\mathrm{d}v} \ ,
\end{equation}
where $v$ is the velocity shift in \kms, $V(v)$ the Stokes V profile, $I(v)$ the Stokes I profile, $c$ is the speed of light, and $\lambda_0$ and $g_0$ the respective average wavelength and Land\'e factor defined earlier.

%------------------------F5
   \begin{figure*}
   {\bf a.} \ \xiA\\
   \includegraphics[width=\textwidth,clip]{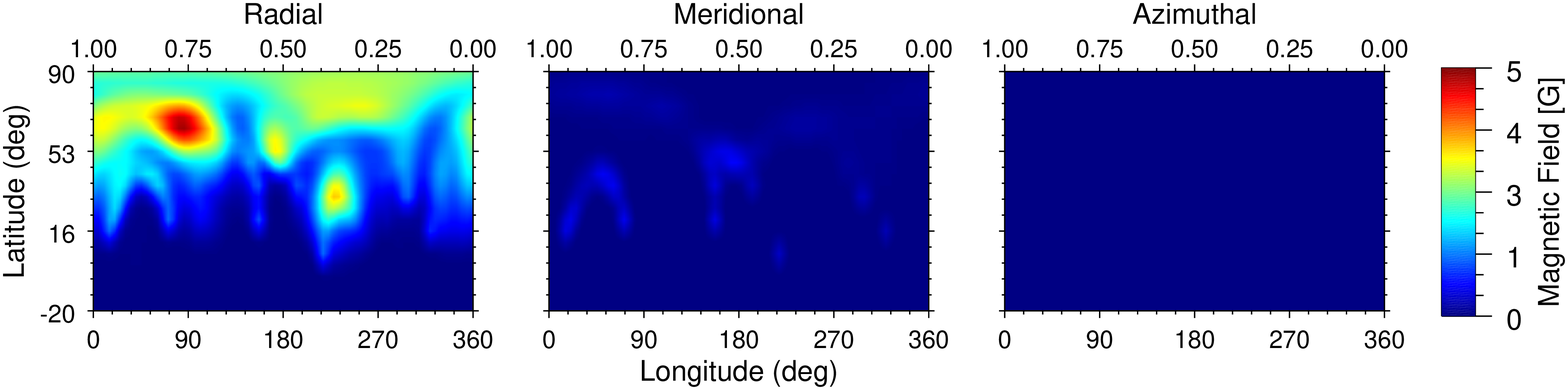}
   {\bf b.} \ \xiB\\
   \includegraphics[width=\textwidth,clip]{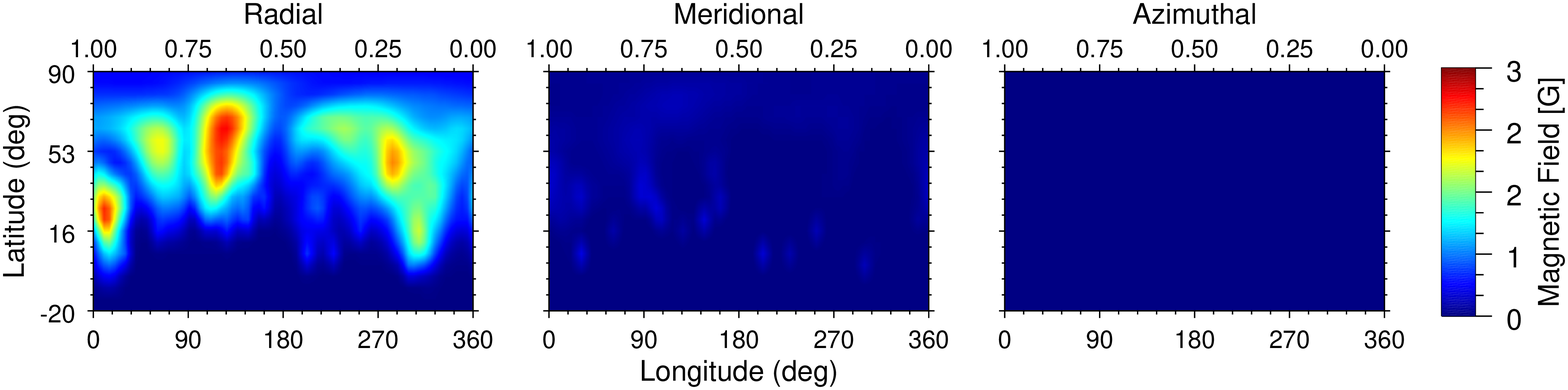}
   \caption{Root-mean-square error maps in Mercator projection from a residual bootstrap analysis. Panel $a$: \xiA\ . Panel $b$: \xiB . From left to right: radial-, meridional-, and azimuthal magnetic field component in Gauss. Regions below a latitude of $-20$\degr\ were set to zero and did not contribute. }
         \label{F5}
   \end{figure*}

\subsection{V-only ZDI solution}\label{S32}

Our ZDI typically proceeds with an alternating minimization between surface temperature and magnetic-field density. That is, we usually start the magnetic inversion with an already pre-iterated temperature image and then alternate all further iterations between temperature and magnetic inversions. However, this is not useful for stars with rotational line broadening below a Doppler resolution threshold, typically the aforementioned five resolution elements across the disk in velocity space. Both our target stars in this paper rotate way too slow to reach this Doppler threshold. In a previous ZDI application, also based on $i$MAP, Strassmeier et al. (\cite{iipeg}) found that the dependence of the temperature inversion on the magnetic inversion is not strong while the dependence of the magnetic inversion on the temperature inversion is strong and leads to a non-trivial scaling of the Stokes~V profile amplitudes and widths. For the present case of two stars with very small $v\sin i$ these dependencies do not reach a numerical meaningful threshold during the iterations. We had done a test with a pre-iterated initial temperature image as the starting image for the Stokes~V inversion of \xiA\ instead of a blank surface with the effective temperature. The outcome was practically identical. Therefore, the missing Stokes~I information during the inversion appears being only a minor issue for our two narrow-lined target stars.

\subsection{Mapping procedure}\label{S33}

A particular feature with $i$MAP is its iterative regularization. This allows a penalty free inversion but with the expense of many iterations and thus CPU time. For the map interpretation it means that we do not indirectly prefer the large-scale surface structure over the small-scale structure by an implicit inversion assumption. The latter is the case for some other ZDI inversions due to their adopted $l^2$ penalty function ($l$ is the angular harmonic expansion coefficient that describes the geometric scale of the magnetic field). In $i$MAP, only the surface gradient vector from one iteration to the next is used for the regularization.

Figure~\ref{F2}ab show the reconstructed sets of Stokes-V profiles for stars A and B, respectively, compared with the SVD-averaged data for all rotational phases. The final solution is only based on Stokes~V. $i$MAP performed a total of about 3000 iterations for the final solution for each star. For each velocity bin, a mean standard error averaged over the velocity domain of 2.85$\times$10$^{-5}$ and 4.53$\times$10$^{-5}$ is achieved for \xiA\ and \xiB, respectively. These values are practically identical to the respective inverse S/N per pixel due to $i$MAP's inversion stopping rule.

Our ZDI Stokes-V maps are shown in orthographic projection in Fig.~\ref{F3}a for \xiA\ and in Fig.~\ref{F3}b for \xiB. The reconstructed total flux density (loosely called field strength) is plotted with a color code also indicating the polarity (red positive and blue negative). Magnetic field strengths are computed as the squared sum of the three vector components. Figure~\ref{F4}a and Fig.~\ref{F4}b show the same magnetic maps but in Mercator-style projection and split into the three vector components.

Figure~\ref{F5}a and Fig.~\ref{F5}b are the root-mean-square (rms) error maps for the three magnetic components for both stars. These maps are from inversions with different initializations for the three field parameters for all surface segments (radial, azimuthal, and meridional flux density) and are run for 100 times on both original data sets. A random generator on the basis of a normal distribution provides the values for each parameter and surface segment. The inversions are run to the same accuracy given by the noise of the data. Note that we set the surface segment values below a stellar latitude of --20\degr\ to zero because its derivatives become too weak to provide any substantial changes during the inversion.

The error map of the radial magnetic field for \xiA\ shows a peak value of 5\,G, the surface average rms error is less than half of this. The rms error for the meridional and azimuthal fields have peak values of around 1\,G and both components have a mean rms error in the sub-Gauss regime. The error map of the radial field for \xiB\ has a peak value of 3\,G, and also sub-Gauss values for the other two field components. Because the obtained errors are nevertheless comparably small indicates that the inversion always settled in or near the same (local) $\chi^2$ minimum and that the final solution is very robust against errors of the initial conditions.
% All error maps exhibit a correlation with the reconstructed field densities. This is expected because we run all inversions without related temperature information.

Our data have a significant phase gap of 0\fp42 due to a snowstorm event at LBT. While we can not do anything against that, we may recall some of the Doppler-imaging tests with artificial data in the literature. For example, Rice \& Strassmeier (\cite{tests}) conducted extensive tests including recoveries with a phase gap of 100\degr (0\fp28) at various phase locations, with moderate and with high S/N data, different inclinations of the stellar rotation axis, etc., and were always capable of basically correctly recovering the original spot locations. In particular also the individual spot contrasts/temperatures and even within the phase gap. These tests also showed that the equatorial regions are more affected than the polar regions for low inclination cases like for $\xi$\,Boo. Also, the higher the S/N the less numerous were the artifacts. The present phase gap is unfortunately larger but so is the S/N of our data. While we do not claim that such a big phase gap has no impact, we are confident that it does not corrupt the image. However, our ZDI images must be considered an approximation.

%------------------------F6 new
   \begin{figure}
   {\bf a.} \ \xiA\\
   \includegraphics[width=8.7cm,clip]{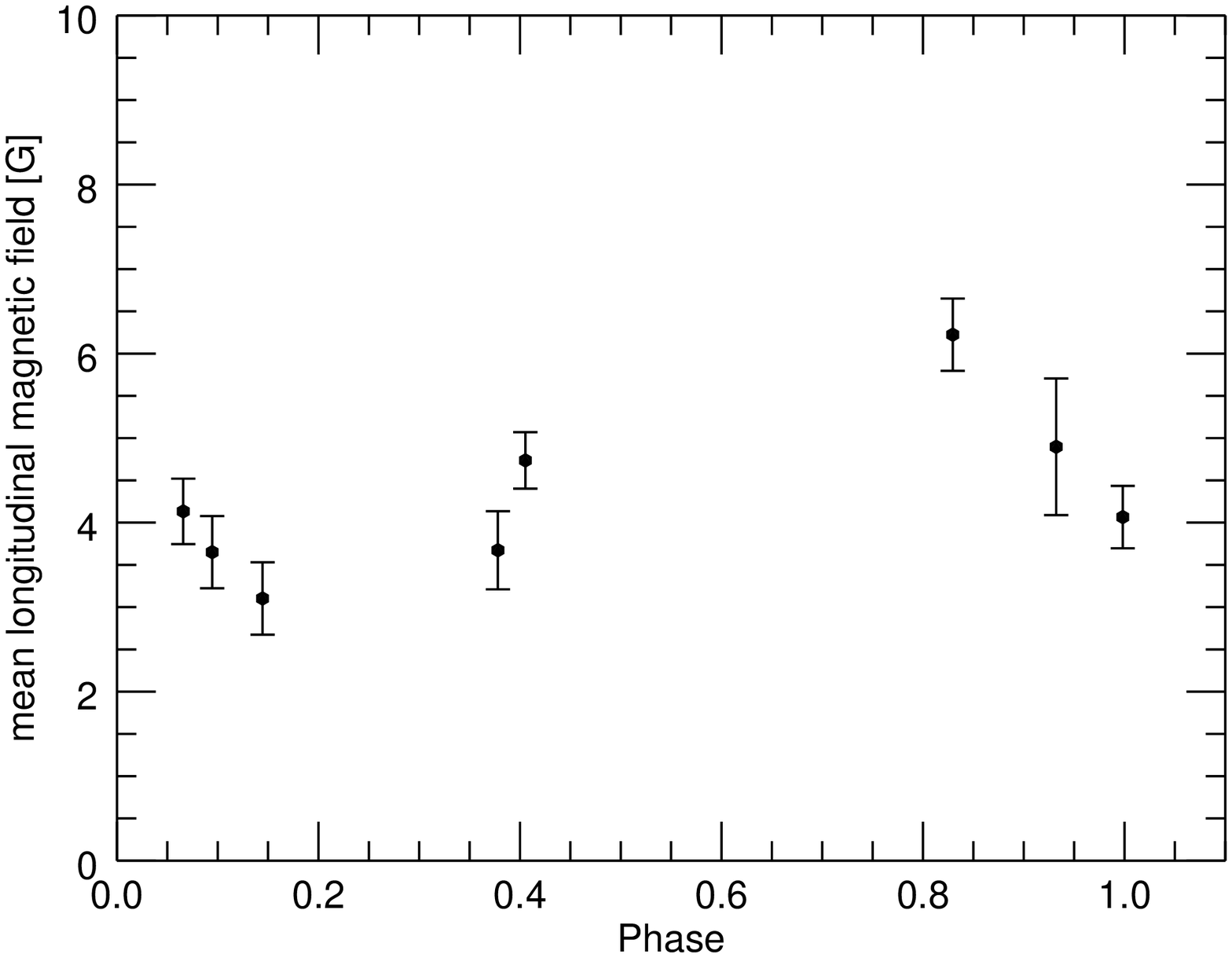}
   {\bf b.} \ \xiB\\
   \includegraphics[width=8.7cm,clip]{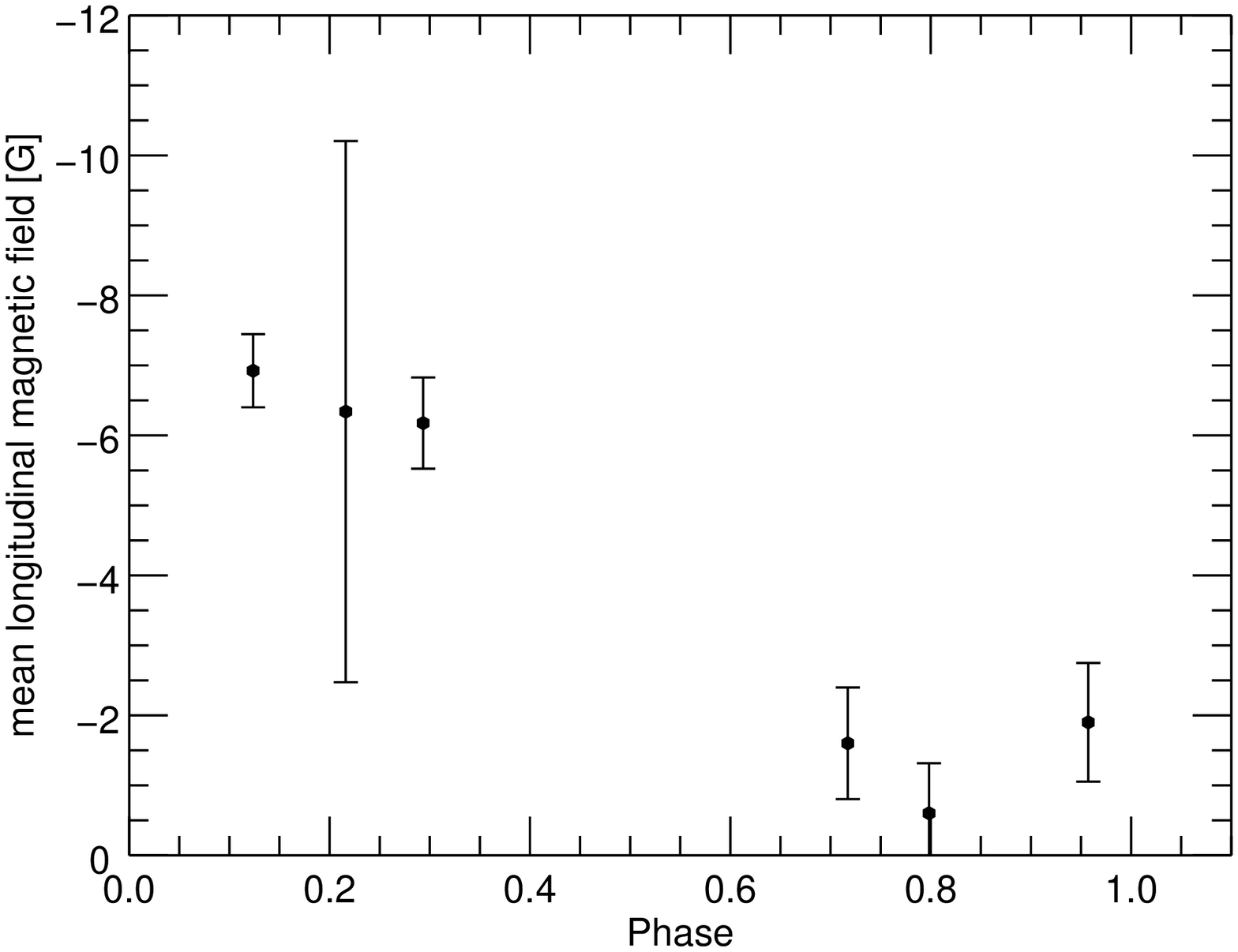}
   \caption{Mean longitudinal magnetic field in Gauss versus rotational phase. Panel $a$: \xiA\ . Panel $b$: \xiB . }
         \label{F6}
   \end{figure}

\subsection{Magnetic surface morphology}\label{S34}

Surface spots are reconstructed with a radial-field density of up to plus/minus 115\,G, a meridional field density of up to plus/minus 30\,G, and an azimuthal field density of up to plus/minus 20\,G for \xiA. Standard errors are up to $\pm$5\,G. For \xiB\ the values are significantly smaller with a radial-field density of up to plus/minus 55\,G, a meridional field density of up to plus/minus 15\,G, and an azimuthal field density of up to plus/minus 15\,G. Its standard errors are up to $\pm$3\,G. Both maps almost satisfy the divergence free condition already after the not-constrained ZDI inversion. By adding only a small amount of a monopole contribution to the maps (1.2 G for \xiA\ and --0.9 G for \xiB) we
obtain a zero integral of $B_{\rm r}$ over the stellar surface.

% ------------------------ Table 2
\begin{table}
  \caption{Magnetic spots on \xiA\ and \xiB\ in May 2019.}\label{T2}
\centering
\begin{tabular}{lllll}
\hline\hline\noalign{\smallskip}
Spot  & Long      & Lat      &  $A$  & $B_{\rm max}$ \\
ID    & (\degr )  & (\degr ) & (\% ) & (G) \\
\noalign{\smallskip}\hline\noalign{\smallskip}
\multicolumn{5}{l}{\xiA :}\\
P & $\approx$180  & +77 &  9.0 & --80 \\
A & 0   & +50 &  5.5 & +90 \\
B & 111 & +43 &  3.8 & +85 \\
C & 235 & +28 &  6.2 & +115 \\
\multicolumn{5}{l}{\xiB :}\\
A & 14  & +19 &  1.7 & +30 \\
B & 60  & +55 &  2.1 & --15 \\
C & 129 & +42 &  4.0 & +30 \\
D & 212 & +38 &  9.5 & --50 \\
E & 311 & +22 &  6.6 & --50 \\
\noalign{\smallskip}\hline
\end{tabular}
\tablefoot{Longitudes and latitudes are given for the approximate spot center. The letter numbering is with increasing longitude. Spot area $A$ is given for the radial field in per cent of the entire surface and for the extent of the magnetic feature down to an arbitrary threshold contrast of (plus/minus)35\,G for \xiA\ and (plus/minus)15\,G for \xiB. $B_{\rm max}$ is the 95 percentile peak radial-field flux density.}
\end{table}

Table~\ref{T2} quantifies the reconstructed spots. \xiA's magnetic morphology is characterized by a very high latitude or even polar spot (dubbed P in Table~\ref{T2}) of negative polarity in combination with three confined low-to-mid latitude spots (dubbed ABC) of positive polarity. A weak meridional component of positive polarity is still reconstructed opposite to the (radial) negative-polarity  near-polar feature. An equally weak azimuthal component of negative polarity is only detected $\approx$180\degr\ away from the meridional feature and at medium latitude. However, neither feature from the meridional and azimuthal components is well constrained, and all are comparably weak with +20$\pm$5~G and --22$\pm$5~G, respectively, thus relatively uncertain. \xiB's magnetic morphology, on the other hand, is dominated by four low-to-mid latitude spots (dubbed ACDE) of mixed polarity, that is two spots with positive and two spots with negative polarity, where the negative and the positive polarity spots each fall within different stellar hemispheres. A fifth feature (dubbed B) of negative polarity is also reconstructed but appears comparably small and weak with just --15\,G and thus remains a bit doubtful. No polar magnetic feature is reconstructed on \xiB. \xiB's meridional and azimuthal components appear morphological comparable to \xiA\ but weaker in strength by a factor two.

Figure~\ref{F6} shows the disk-averaged longitudinal field versus rotational phase. The low phase-averaged values of +4.5\,G for \xiA\ and --5.0\,G for \xiB\ indicate an almost even distribution of polarities. Overall, \xiA's disk maintains an apparent net positive field while \xiB's disk a net negative field, albeit both very weak. The respective field dispersions are $\pm$1.5\,G and $\pm$3\,G. Our value for \xiA\ in 2019 is at the lower end of the 4.1--11.2\,G range measured in Morgenthaler et al. (\cite{morgen}) for the years 2007--2011, but apparently in good agreement with the years 2008 and 2010 for which longitudinal fields of +4.6$\pm$3.1\,G and +4.1$\pm$5.3\,G, respectively, were given. Plachinda \& Tarasova (\cite{pla:tar}) had obtained values in the range --10 to +30\,G, mainly from 1998, and Hubrig et al. (\cite{hub}) in the range --15 to +50\,G, using the same Crimean Observatory Stokesmeter at $R\approx 30,000$ but in 1990. These two sets are thus of much lower spectral resolution. Our mean value for \xiB\ in 2019 is more than three times smaller than the snapshot value in Marsden et al. (\cite{marsden}) of --18.9$\pm$0.5\,G in 2012, but of same (negative) sign.
Both star's maps are dominated by the radial-field component. Neither shows a strong azimuthal component as reconstructed by Petit et al. (\cite{petit}) and Morgenthaler et al. (\cite{morgen}) for most of their maps for \xiA, in particular for the epochs in 2007, 2009, 2010, and 2011 (four out of seven maps). For \xiB\ no other maps exist and thus no comparison is possible. Regarding field strengths, our reconstructed map for \xiA\ has a spot-averaged field strength of $\approx$100~G compared to the range of 30--70~G for the five maps in Morgenthaler et al. (\cite{morgen}). We recall though that the Morgenthaler et al. maps were for epochs between 2007 to 2011, while our map is from 2019. Any direct comparison of field strengths is thus difficult, mostly because the surface spots likely had changed during that time. Nevertheless, our radial flux-density values are up to a factor of two higher than the previously published values by Petit et al. (\cite{petit}) and Morgenthaler et al. (\cite{morgen}).

The total magnetic energy is with 86\%\ for \xiA, and 89\%\ for \xiB, almost exclusively in the radial component while the meridional (13\%\ for \xiA\ and 10\%\ for \xiB) and azimuthal ($\approx$1\%\ for \xiA\ and \xiB) components are comparably weak. The (radial) field distribution is 69\%\ axisymmetric and 31\%\ non-axisymmetric for \xiA , and 57\%\ axisymmetric and 43\%\ non-axisymmetric for \xiB. We used a scalar spherical harmonic decomposition of each resulting individual map (radial, meridional, and azimuthal) independently. This orthonormal decomposition is used exclusively to determine the symmetry properties of the individual components and the energy per spherical harmonic degree.

\section{Summary and outlook}\label{S4}

In this paper, we present new CP spectra for both components of the visual binary $\xi$\,Boo~AB (G8V+K5V) and use them for a detailed analysis of their magnetic surfaces. Our new spectra are of currently highest possible quality with a spectral resolution of 130\,000 and S/N per pixel of up to 2100 for \xiA\ and 1350 for \xiB. It enables a line-by-line detection of the Zeeman pattern even without line averaging. Together with the full radiative-transfer treatment of the local line profile in our ZDI inversion, we arrive at CP line-profile fits with spatially resolved magnetic-field densities of only few per cent errors.

Both stars still hold some surprises. The magnetic-field polarities on the warmer G8V star \xiA\ appear systematically separated in latitude (negative polarity around the visible polar region, positive polarity around the equator) while on the cooler K5V star the field appears separated in longitude (positive polarity dominantly in the leading hemisphere, negative polarity in the trailing hemisphere). No polar magnetic fields are reconstructed on \xiB. The morphology on both stars is dominated by the radial component of the field vector with magnetic energies at the 86\%\ level for \xiA\ and 89\%\ for \xiB. Its reconstructed peak densities are plus/minus 115$\pm$5\,G for \xiA\ and plus/minus 55$\pm$3\,G for \xiB. The value for \xiA\ is approximately twice as large than those from the previous ZDI reconstructions by Petit et al. (\cite{petit}) and Morgenthaler et al. (\cite{morgen}) while the peak meridional and azimuthal densities of ($\pm$)20--30~G are a factor of two smaller than the previous reconstructions. Ros\'en et al. (\cite{rosen15}) argued that cross-talk between the radial and meridional field components can occur when only circular polarization is used in the magnetic inversion. This may be because Stokes~V is formally only sensitive to the line-of-sight component of the magnetic field. In Stokes~IV maps the radial component appeared always to be the strongest of the three while it is the weakest in the IQUV inversion of Ros\'en et al. (\cite{rosen15}). We can not decide whether this discrepancy is due to the unaccounted Stokes I and/or linear polarization (LP) or, for example, related to the $l^2$ penalty function applied in the inversion by Ros\'en et al. (\cite{rosen15}). In any case, one would still need a physical effect that explains why the field density of a field parallel to the surface (meridional or azimuthal) would be repeatedly higher than the radial field. Explanations for such a dominant horizontal field were proposed in ZDI studies and in numerical 3D magneto-hydrodynamic simulations. For example, in the simulations presented by Brown et al. (\cite{brown2010}) they found striking wreaths of magnetism in the midst of the convection zone, with a toroidal magnetic field maintained by the differential rotation, and with a poloidal field from turbulent correlations between the convective flows and magnetic fields. Such magnetic-field structure may eventually, at least partly, propagate up to the surface and could then be seen.

Among the next steps will be adding LP to the line-profile inversion of $\xi$\,Boo. Unfortunately, LP in spectral lines of the Sun is not only typically up to ten times weaker than CP (e.g., Stenflo \cite{stenflo89}) but also much more complex and divergent in its line formation (Landi Degl'Innocenti \& Landolfi \cite{landi}, Sampoorna et al. \cite{samp}), and thus much more uncertain. The LP data of $\xi$\,Boo will eventually serve this purpose in the future.

\begin{acknowledgements}
It is a pleasure to thank the German Federal Ministry (BMBF) for the year-long support for the construction of PEPSI through their Verbundforschung grants 05AL2BA1/3 and 05A08BAC as well as the State of Brandenburg for the continuing support of AIP and the LBT (see https://pepsi.aip.de/). We also thank an anonymous referee for the many helpful comments and suggestions. LBT Corporation partners are the University of Arizona on behalf of the Arizona university system; Istituto Nazionale di Astrofisica, Italy; LBT Beteiligungsgesellschaft, Germany, representing the Max-Planck Society, the Leibniz-Institute for Astrophysics Potsdam (AIP), and Heidelberg University; the Ohio State University; and the Research Corporation, on behalf of the University of Notre Dame, University of Minnesota and University of Virginia. This work has made use of the VALD database, operated at Uppsala University, the Institute of Astronomy RAS in Moscow, and the University of Vienna. This research has made use of NASA's Astrophysics Data System and of CDS's Simbad database which we both gracefully acknowledge.
\end{acknowledgements}

\begin{appendix}

\section{Observing log}

% ------------------------ TA1
\begin{table*}[!tbh]
\caption{Observing log.}\label{TA1}
\begin{tabular}{llllllll}
\hline\hline\noalign{\smallskip}
 TCB mid & $\phi$ & $t_{\rm exp}$ & $\Delta\lambda$ (blue; red) & \multicolumn{2}{c}{S/N$_{\rm V}$} & \multicolumn{2}{c}{S/N$_{\rm I}$}\\
 (+245\dots)  & (Eqs.\,1) & (min)& (\AA )                   &  blue & red & blue & red \\
\noalign{\smallskip}\hline\noalign{\smallskip}
\xiA : & & & & & & & \\
 8609.7427427     & 0.829 & 5  &  4800-5441; 6278-7419 & 1405 & 1749 & 2145 & 2982 \\
 8610.8282148     & 0.998 & 5  &  4800-5441; 6278-7419 & 1368 & 1741 & 2016 & 2910 \\
 8611.7681903     & 0.144 & 5  &  4800-5441; 6278-7419 & 1191 & 1594 & 1591 & 2392 \\
 8616.8333173     & 0.931 & 5  &  4800-5441; 6278-7419 &  567 &  820 & 1173 & 1773 \\
 8617.6923984     & 0.065 & 5  &  4800-5441; 6278-7419 & 1390 & 1739 & 2118 & 2848 \\
 8617.8769492     & 0.094 & 5  &  4800-5441; 6278-7419 & 1699 & 2128 & 2448 & 3383 \\
 8619.7004465     & 0.377 & 5  &  4800-5441; 6278-7419 & 1284 & 1685 & 1969 & 2804 \\
 8619.8758547     & 0.405 & 5  &  4800-5441; 6278-7419 & 1539 & 1976 & 2369 & 3363 \\
\xiB : & & & & & & & \\
 8609.7860453     & 0.123 & 10 &  4800-5441; 6278-7419 &  956 & 1351 &  994 & 1848 \\
 8610.8816143$^a$ & 0.216 & 10 &  4800-5441; 6278-7419 &  789 &   85 &  297 &  236 \\
 8611.8163769     & 0.293 & 10 &  4800-5441; 6278-7419 &  787 & 1171 &  893 & 1718 \\
 8616.8730350     & 0.717 & 10 &  4800-5441; 6278-7419 &  818 & 1187 &  867 & 1713 \\
 8617.8435912$^b$ & 0.798 & 5  &  4800-5441; 6278-7419 &  605 &  954 &  705 & 1472 \\
 8619.7390032     & 0.957 & 10 &  4800-5441; 6278-7419 &  510 &  796 &  556 & 1084 \\
 \noalign{\smallskip}\hline
\end{tabular}
\tablefoot{The first column gives the Barycentric Coordinate Time (TCB) for the time of mid exposure for Stokes I. The second column is the rotational phase based on the respective ephemeris in Eq.~(\ref{eq1}) for Stokes~I. S/N is per pixel and is the 95\%\ quantile within the respective wavelength region $\Delta\lambda$. $a$. Bad weather. $b$. Exposure time was accidentally set to 4\,min in the blue, and to 5\,min in the red. }
\end{table*}

\end{appendix}

\end{document}